\documentstyle[prl,twocolumn,aps,epsf,epsfig,amsmath,amssymb]{revtex}
\begin{document}

\twocolumn[
\hsize\textwidth\columnwidth\hsize\csname@twocolumnfalse\endcsname
\draft

\title{Temperature dependent weak field Hall resistance in 2D carrier
  systems}
\author{S. Das Sarma and E. H. Hwang}
\address{Condensed Matter Theory Center, 
Department of Physics, University of Maryland, College Park,
Maryland  20742-4111 } 
\date{\today}
\maketitle

\begin{abstract}

Using the Drude-Boltzmann semiclassical transport theory, we calculate
the weak-field Hall resistance of a two-dimensional system at low
densities and temperatures, assuming carrier scattering by screened
random charged impurity centers. The temperature dependent 2D Hall
coefficient shows striking non-monotonicity in strongly screened
systems, and in particular, we qualitatively explain the recent
puzzling experimental observation of a decreasing Hall resistance with
increasing temperature in a dilute 2D hole system. We predict that the
impurity scattering limited Hall coefficient will eventually increase
with temperature at higher temperatures.


\end{abstract}
\vspace{0.5cm}
]

\newpage

The behavior and the properties of the apparent two dimensional (2D)
``metallic'' phase continue to attract substantial attention
\cite{review} 
from experimentalists and theorists alike, even a decade after its
original discovery \cite{Krav94}. In particular, the original
observations on the 
strong metallic (i.e. $d\rho/dT >0$) temperature dependence of the
2D resistivity, $\rho(T)$, where the resistivity may increase
by as much as a factor of $3-4$ for a modest increase in
temperature (e.g. $T=100 mK - 3K$) were followed by intriguing
observations of large  magnetoresistance
in an applied parallel magnetic
field. Phenomenologically the observed ``metallicity'', defined as the
maximum temperature induced enhancement of $\rho(T)$, exhibits
strong system dependence, with 2D p-GaAs hole system being the most
metallic and 2D n-GaAs electron system being the least metallic with
the 2D Si-based electron systems having intermediate metallicity. This
system-dependent variation can be understood
on the basis of our theoretical prediction \cite{DH2004} that the
metallicity arises from 2D screening properties, and is therefore
controlled in the zeroth order theoretical prescription by the
dimensionless parameters $q_{TF}/2k_F$ and $T/T_F$, where $q_{TF}$,
$k_F$, $T_F$ are respectively 2D Thomas-Fermi 
screening wave vector, the Fermi wave vector, and the Fermi
temperature \cite{Ando_rmp}. Since $q_{TF}/2k_F \propto m n^{-1/2}$
and $T/T_F 
\propto mn^{-1}$ in 2D, the metallicity increases with increasing
(decreasing) carrier effective mass $m$ (carrier density $n$). This is
why metallicity is stronger at lower carrier densities and/or in
higher effective mass semiconductor systems.

The idea \cite{DH1} of the
strongly temperature dependent screened charged impurity 
effective disorder
being the qualitative reason underlying the striking metallic behavior
of dilute 2D carrier systems has come to be known as the ``screening
theory'' since the screening-induced regularization of the bare
Coulombic impurity disorder (in contrast to zero-range white noise
disorder) intrinsic to semiconductor systems is the crucial physical
mechanism in this theory. The screening theory has been applied, with
reasonable qualitative success, to explain the temperature and density
\cite{DH2004,DH1,lilly} dependence as well the parallel magnetic field
dependence \cite{gold_b}
of the 2D ``metallic'' resistivity \cite{DH2004,DH1}.
Motivated by a puzzling recent experimental observation \cite{gao}, we
develop 
in this Letter the first theory, based on the screening model, for the
temperature dependence of the weak-field Hall resistance in dilute 2D
carrier systems. Our theory is in excellent qualitative agreement with
the experimental results \cite{gao}
on the temperature dependence of the weak
field Hall resistance although quantitative discrepancies remain.

The recent Hall resistance measurements \cite{gao} take on particular
significance since the experimental data reported by Gao {\it et al.}
\cite{gao} disagree {\it qualitatively} with the so-called ``interaction
theory'' \cite{zala,zala2}, which has recently been much discussed and
debated 
in the literature
\cite{zala,zala2,gold2003,inter1,inter2,inter3,inter4,DH2004}. The
interaction theory complements the 
screening theory by carrying out a perturbative diagrammatic
calculation to include {\it all} (i.e. not just screening) interaction
corrections to the 2D conductivity in the weak-disorder (the so-called
``ballistic'' regime), low-temperature limit defined by
($\hbar=k_B=1$) $\tau_0^{-1} \ll T \ll T_F$, where $\tau_0$ is the
$T=0$ transport (``Drude'') relaxation time, $\sigma_0\equiv
\sigma(T=0) =ne^2\tau_0/m$. The interaction theory, which purportedly
improves and extends the screening theory by including higher-order
interaction corrections, has several
limitations: (1) the theory is restricted to only
small temperature induced corrections $\delta \sigma(T)$ to $\sigma_0$
with $\sigma(T) = \sigma_0 + \delta \sigma(T)$, and $|\delta \sigma|
\ll \sigma_0$ -- as such this theory is, by construction, incapable of
explaining the large temperature-induced changes
in the conductivity observed
in the 2D ``metallic'' phase; (2) the theory is necessarily restricted
to very low temperatures $T \ll T_F$, which may not be achieved
experimentally; (3) the theory has only been developed for 2D
systems with white-noise bare impurity disorder, i.e. for an
unrealistic model of zero-range bare impurity scattering potential (in
reality, the impurities in 2D semiconductor structures are random
Coulombic charge centers, not zero-range neutral scatterers); (4) the
interaction theory predicts only the leading-order temperature
dependence (``linear-in T'') for 2D conductivity as $\delta \sigma(T)
= \sigma_0 C_1 (T/T_F)$, where the temperature-independent coefficient
$C_1$ is a universal (but unknown) function of density defined by the
Fermi liquid ``triplet'' interaction parameter $F_0^{\sigma}$.
Depending on the value of
$F_0^{\sigma}$, $\delta \sigma(T)$ could be positive or negative.
In spite of the fact
that the only real prediction of the interaction theory for $\sigma
(T)$ is that the leading-order thermal correction to the 2D
conductivity is linear in $T$ (with an unknown positive or negative
coefficient), and the experimental $\sigma(T)$ is rarely linear over any
appreciable temperature range, there have been experimental
attempts \cite{inter1,inter2,inter3,inter4} to attribute the 2D
metallicity to interaction effects by 
fitting the experimental $\delta \sigma(T) \equiv \sigma(T) -
\sigma_0$, where $\sigma_0$ is obtained by a linear extrapolation to
$T=0$, to the
formula $\delta \sigma(T)/\sigma_0 = C_1 (T/T_F)$ and thereby obtain
the fitted coefficient $C_1$ and consequently  extract the Fermi
liquid parameter $F_0^{\sigma}$. It may be appropriate here to mention
that the screening theory also predicts a leading-order linear
temperature correction to $\sigma(T)$.
While both the screening theory and the interaction
theory predict $\delta \sigma(T) \sim T/T_F$, the screening theory is
not limited to just the leading-order temperature dependence and can
be applied \cite{DH2004,DH1,lilly}to calculate the complete
$\sigma(T)$ -- in fact, 
the screening theory becomes more accurate at higher temperatures. The
screening theory is, however, a self-consistent field theory which
only includes the effects of screened effective disorder arising from
the charged impurity scattering leaving out all higher-order
(i.e. beyond screening) effects of interaction. Therefore, the
applicability of the screening theory at very low-density strongly
interacting 2D system is suspect and is only of qualitative
validity. The importance of the interaction theory \cite{zala,zala2},
in spite of its limitations, arises from its very general and
universal nature where the temperature correction to the 2D
conductivity is linked to universal functions of Fermi liquid
interaction parameters.

Since the temperature-dependent conductivity itself (at least, the
leading-order temperature correction) cannot distinguish between the
interaction and the screening theory (with both predictions being
$\delta \sigma/\sigma_0 \propto T/T_F$), it becomes imperative to look
for other more definitive signatures for interaction effects in
low-density and low-temperature 2D transport properties. This is where
the temperature dependence of weak-field 2D Hall resistivity
$\rho_H(T)$ takes on great significance as was emphasized
in ref. \onlinecite{zala2}, and recently, in
ref. \onlinecite{gao}. The interaction theory makes \cite{zala2}
very specific (and falsifiable) predictions about the connection
between the (leading-order) temperature corrections to 2D conductivity
$\delta \sigma(T)$ and Hall resistivity $\delta \rho_H(T)$ where
$\rho_H(T) = \rho_H^0 + \delta \rho_H(T)$ with $\rho_H^0 \equiv \rho_H
(T\rightarrow 0)$. The predicted
connection between $\delta \sigma(T)$ and $\delta \rho_H(T)$, which
has been discussed in detail with exemplary clarity in
refs. \onlinecite{gao} and \onlinecite{zala2},
arises from the fact that in the interaction theory both of
these temperature corrections are controlled by the same Fermi liquid
parameters, and as such, a detailed and careful measurement of
$\sigma(T)$ at low temperatures necessarily {\it completely}
determines the low-temperature behavior of $\rho_H(T)$ in the same
sample. Gao {\it et al}. carried out \cite{gao} such a
comparison between $\rho_H(T)$ and $\sigma(T)$ using the interaction
theory \cite{zala2} 
in an extremely high-quality 2D p-GaAs hole system which
is metallic at extraordinary low densities down to $10^{10} cm^{-2}$
(where the 2D hole system should be very strongly interacting, making
it a perfect system for testing the internal consistency of the
interaction theory). The outcome as detailed in ref. \onlinecite{gao} is a
spectacular {\it qualitative} failure of the interaction theory: In the
density range studied \cite{gao} by Gao {\it et al}. the Fermi liquid
parameters extracted from their experimental $\sigma(T)$ data should
lead to, according to the interaction theory results of
ref. \onlinecite{zala2}, 
essentially a constant temperature-independent $\rho_H(T)$ except at
the lowest temperature the interaction theory predicts a small (less
than 1\%) increase in $\rho_H(T)$ with increasing temperature, whereas
Gao {\it et al}. find a smooth {\it decrease} ($\sim 20 \%$) in
$\rho_H(T)$ with increasing temperature as $T$ varies from
100 mK to 1K. Thus the interaction theory  disagrees with the
temperature dependent Hall resistance data of ref. \onlinecite{gao},
both qualitatively and quantitatively.

In this Letter we present the first calculated results for the 2D
temperature dependent Hall resistance using the screening theory
formalism. We solve the semiclassical Drude-Boltzmann transport
equation \cite{phonon} for the weak-field Hall resistance finding that
the 2D 
Hall resistivity $\rho_H$ can be written as ($c=1$)
\begin{equation}
\rho_H = \frac{B}{ne}r_H,
\end{equation}
where $B$ is the applied weak magnetic field 
and $r_H$, the so-called Hall ratio [4], is given by
\begin{equation}
r_H = {\langle \tau^2 \rangle}/{\langle \tau \rangle^2},
\end{equation}
where $\tau$ is the (energy and temperature dependent) 2D carrier
transport scattering time (the so-called momentum relaxation time)
determined by the {\it screened} charged impurity scattering
\cite{DH2004}, and 
$\langle \tau \rangle$ is the thermal average over the finite
temperature Fermi distribution function -- the detailed definitions of
$\tau$ and $\langle \tau \rangle$ are given in the Appendix of
ref. \onlinecite{DH2004} and will not be repeated here.
We note that $\langle \tau \rangle$ determines the Drude-Boltzmann
conductivity through the relation $\sigma(T) = ne^2 \langle \tau
\rangle /m$.

At $T=0$, $\langle \tau^2 \rangle \equiv \langle \tau \rangle^2$,
giving $r_H =1$ so that $\rho_H(T=0) \equiv \rho_H^0 = B/ne$, the
classical Hall formula. At finite temperatures it is well known that
the Hall ratio $r_H \neq 1$ due to thermal corrections, and below we
present our results for the calculated temperature-dependent Hall
resistance in the Drude-Boltzmann theory of screened charged impurity
scattering. Following the standard notation we define the Hall
coefficient $R_H = d\rho_H/dB = r_H/ne$.

Our numerical results (shown below) for the temperature dependent Hall
resistance are calculated assuming only carrier scattering by screened
charged impurity scattering, leaving out phonon scattering 
(except in the inset of Fig. 2(a)). For
screening the Coulombic disorder potential from the unintentional
random charged impurities in the background and interfaces we use the
random phase 
approximation (RPA) augmented by the Hubbard approximation (HA) for
the local field correction \cite{drag,ha}. It is well-known that HA
quantitatively improves upon the RPA, particularly at the very low
carrier densities of interest \cite{gao}, by including some short-range
exchange-correlation effects beyond the self-consistent field
approximation of RPA. Our calculations are done for the realistic
GaAs/Ga$_{1-x}$Al$_x$As quantum well systems used in ref. 9 with the
finite-width form-factor effects \cite{DH2004,Ando_rmp} associated
with the quantum 
well confinement included in the theory. The charged impurities are
assumed to be uniformly randomly distributed in the background and
interfaces, which set the overall resistivity scale
(i.e. $\sigma_0^{-1}$) in the system. All calculations are done for a
quantum 
well width of 100\AA \;  corresponding to the sample of
ref. \onlinecite{gao}.

In Fig. 1 we show our calculated conductivity $\sigma(T)$ and Hall
coefficient $R_H(T)$ (inset) as a function of temperature for several hole
densities. Both $\sigma(T)$ and $R_H(T)$, in
agreement with the experimental data \cite{gao}, decrease with increasing
temperature, exhibiting some interesting non-monotonicity that we
discuss below. The overall decrease for $\sigma(T)$, about a
factor of $3-4$, and for $R_H(T)$, about 10 -- 20\%, are in excellent
agreement with experiment although there are quantitative
discrepancies in the details, which is to be expected given the highly
approximate nature of our theory.

The most significant quantitative discrepancy between experiment
and theory is that the overall temperature scale for
the theory is somewhat wider than that in the experiment \cite{gao}. 
While this problem can be somewhat rectified by
including additional scattering mechanisms (e.g. phonons, surface
roughness, alloy disorder, remote impurities) which are invariably
present in real 2D semiconductor systems, our goal in this paper is to
avoid excessive data fitting (which would not be particularly
meaningful from the perspective of fundamental understanding of the 2D
metallic phase) in order to establish a basic zeroth order qualitative
understanding of the temperature dependent 2D Hall effect.  We
therefore accept the quantitative discrepancy as the signature of
the approximate nature of our theory, emphasizing the fact that the
theory seems to be an excellent qualitative description of the
experimental observations in ref. \onlinecite{gao}.

In Fig. 2 we show our calculated Hall ratio $r_H(T)$ and the Hall
coefficient, $R_H(T)$, as a function of temperature. In particular,
for $R_H(T)$ we carry out a direct comparison between our screening
theory and the experimental results from ref. \onlinecite{gao}
(cf. Fig. 2(b) in 
ref. \onlinecite{gao}). For the sake of comparison the interaction
theory result \cite{zala2}
for $R_H(T)$, adopted from Fig. 2(b) of ref. \onlinecite{gao}, is also
shown 
on the same plot. First, we note that while the interaction theory
disagrees qualitatively and radically with the experimental results,
our screening theory is in very good qualitative agreement. The
screening theory catches well


\begin{figure}
\epsfysize=2.5in
\centerline{\epsffile{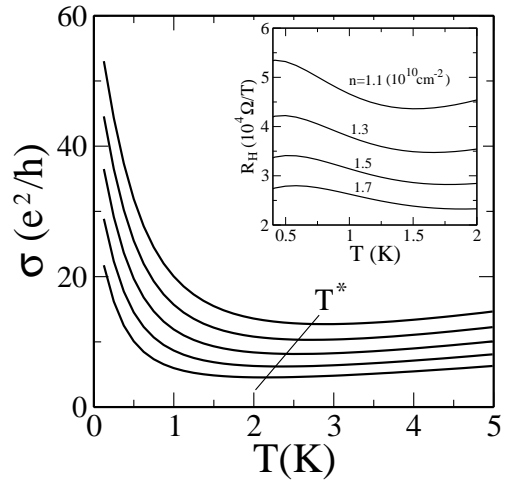}}
\vspace{0.5cm}
\caption{Calculated impurity scattering limited conductivity $\sigma(T)$ 
as a function of temperature for different densities
  $n=1.1$, 1.3. 1.5, 1.7, 1.9$\times 10^{10}cm^{-2}$ (bottom to
  top). $T^*$ indicates the
  temperature at which $d\sigma/dT$ changes sign.
Inset shows the calculated Hall coefficient $R_H(T)$ for different densities. 
}
\end{figure}


\noindent
the overall magnitude ($\sim 20\%$) of the
temperature induced {\it decrease} in the Hall coefficient although
there are  quantitative discrepancies. The most important
discrepancy is that the temperature scale for the temperature
dependence is off by about 400 mK in the sense that the experimental
temperature scale (the bottom abscissa in Fig. 2(b)) goes from 0 to
1.5K whereas our theoretical scale (the top abscissa of Fig. 2(b))
goes from 0.5K to 2.0K. This same discrepancy can be seen in Fig. 1
also where $T^*$, the temperature scale where $d\sigma/dT$ changes sign
at finite temperatures (the so-called ``quantum-classical crossover''
\cite{DH1} phenomenon), is consistently higher by roughly a factor 3 in 
the theory (compare, for example, $T^*$ in our
Fig. 1(a) with the corresponding experimental $T^*$ in Fig. 1 of
ref. \onlinecite{gao}). Currently, we have no good explanation for
this discrepancy 
in the actual value of $T^*$ --- it could, for example, be arising from
an enhanced effective mass (e.g. due to many-body electron-electron
interaction) or due to the effects of other scattering mechanisms such
as remote impurities, which is known to
suppress the theoretical $T^*$. For our qualitative theory of
the temperature dependent Hall effect we just accept this discrepancy
in the temperature scale and shift our theoretical results by 0.5 K in
Fig. 2(b), getting excellent qualitative agreement between experiment
\cite{gao} and the screening theory.

In discussing the overall temperature dependence of the Hall ratio
$r_H$ shown in Fig. 2 (a), where $R_H \equiv r_H/(ne)$, we see that
there is a very striking non-monotonicity in $r_H$ over the temperature
scale $0-4$K which has not been observed experimentally. But, the
experimental results of ref. \onlinecite{gao} clearly suggest
tantalizing signs of 
possible non-monotonic behavior both at the low 


\begin{figure}
\epsfysize=2.2in
\centerline{\epsffile{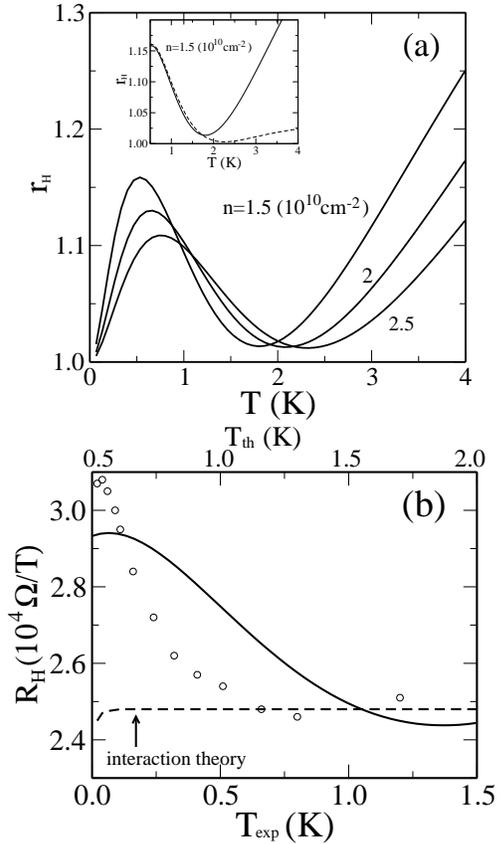}}
\epsfysize=2.2in
\centerline{\epsffile{fig2b.eps}}
\vspace{0.5cm}
\caption{(a) Calculated Hall ratio $r_H(T)$ as a function of
  temperature for different densities by considering only screened charged
  impurities. The phonon contribution (dashed line) is given in the
  inset for $n=1.5 \times 10^{10}cm^{-2}$.
(b) Calculated Hall coefficient  with experimental data [8] and the
  result of the interaction
  theory for hole density $n=1.65\times  10^{10} cm^{-2}$. 
  The scale of $R_H$ is set by experiment [8].
}
\end{figure}


\noindent
and the high
temperature ends with the data in ref. \onlinecite{gao} unfortunately
ending (both 
at low and high temperatures) precisely where the non-monotonicity may
just be appearing. Since the temperature scale of our
screening theory is off by about 500 mK, the non-monotonicity at the
low temperature end (i.e. the initial rise of $r_H$ with increasing
$T$) is probably not observable because carrier heating is likely to
prevent real low carrier temperatures from being  achieved. The higher
temperature rise of $r_H(T)$, beyond $T>2K$ in Fig. 2(a), is a true
prediction of our theory which should be experimentally tested, but
this eventual increase of $r_H(T)$ with $T$ is likely to be masked and
considerably suppressed by phonon scattering effects which become
qualitatively important in low-density p-GaAs 2D hole systems of
interest here already at $T \ge 1K$ (see the inset of Fig. 2(a)). 
Our preliminary estimate of
phonon scattering effects on the temperature dependent Hall
resistivity (the results shown in Figs. 1 and 2, we emphasize, include
only carrier scattering by background screened charged impurities)
suggests that the maximum value of $r_H$ is unlikely to exceed 1.1,
and therefore the size of the non-monotonicity 
is suppressed as shown in the inset of Fig. 2(a).

Before concluding, we point out that, according to our screening
theory, the decreasing $R_H(T)$ with increasing temperature observed
in Ref. \onlinecite{gao} and manifest in our Figs. 1 and 2 is
obviously {\it not} 
an asymptotic leading-order temperature behavior, and as such, it is
missed entirely in the interaction theory which predicts a
qualitatively different behavior. We can calculate the analytic
leading order temperature dependence of the Hall ratio $r_H(T)$ in the
screening theory and find it to be $r_H(T) = 1 +
\frac{\pi^2}{3}(\frac{T}{T_F})^2$, i.e. $\delta \rho_H(T) \sim T^2$,
in the strictly 2D limit. This is different from the
leading-order prediction of the interaction theory, but this is
understandable since the origin of the temperature correction
in $R_H$ is
different in the two theories: in the screening
theory it is a Fermi surface averaging effect at finite temperatures
whereas in the interaction theory, it is a true many-body Fermi liquid
renormalization effect. Our results,
as compared with experimental data of ref. \cite{gao}, demonstrate
that the asymptotic leading-order theoretical temperature dependence
is not meaningful in understanding the temperature dependence of the
2D Hall coefficient --- the temperature dependence of $R_H(T)$ arises
from non-asymptotic higher-order temperature effects and the asymptotic
leading-order temperature dependence remains inaccessible
experimentally because it happens at extremely low temperatures. We
have argued elsewhere \cite{DH2004} that the same may be true for the
temperature 
dependence of the 2D longitudinal resistivity also, 
which is rarely linear in the experimental data.

We  have developed the screening theory for
the temperature dependent Hall effect in 2D ``metallic'' systems. Our
results are in reasonable qualitative  agreement
with the recent experimental observations of Gao {\it et al.}
\cite{gao}. We 
explain why the experimental data \cite{gao} disagree qualitatively with
the interaction theory \cite{zala2} by suggesting that the asymptotic
leading-order temperature dependence  of the interaction theory
is simply inaccessible experimentally and is therefore physically
uninteresting. We predict that the Hall coefficient, which decreases
\cite{gao} 
with increasing temperature upto the highest measurement temperature
($\sim 1.2$K), will eventually increase by $5-10$\% when
the temperature is increased to 2K or beyond, showing an interesting
non-monotonic behavior.

This work is supported by US-ONR, NSF, and LPS.



\begin{thebibliography}{99}

\bibitem{review} S. Das Sarma and Hwang, cond-mat/0411528 (2004);
S. V. Kravchenko and M. P. Sarachik, Rep. Prog. Phys. {\bf 67}, 1
(2004).

\bibitem{Krav94}S. V. Kravchenko {\it et al},
  \prb {\bf 50}, 8039 (1994).

\bibitem{DH2004} S. Das Sarma and E. H. Hwang, Phys. Rev. B {\bf 69},
  195305 (2004). 


\bibitem{Ando_rmp} T. Ando {\it et al}., \rmp {\bf 54} 437, (1982).


\bibitem{DH1} S. Das Sarma and E. H. Hwang, \prl {\bf 83}, 164 (1999);
  \prb {\bf 61}, R7838
  (2000).

\bibitem{lilly} M. P. Lilly {\it et al}., Phys. Rev. Lett. {\bf 90},
  056806 (2003); V. Senz {\it et al}., \prl
  {\bf 85}, 4357 (2000); A. Lewalle {\it et al}., \prb {\bf 66}, 075324
  (2002); S. Das Sarma and E. H. Hwang, \prl {\bf 93}, 269703 (2004).

\bibitem{gold_b}V. T. Dolgopolov and A. Gold, JETP Lett. {\bf 71}, 27
  (2000).

\bibitem{gao} X. P. A. Gao {\it et al}., Phys. Rev. Lett. {\bf 93},
  256402 (2004).

\bibitem{zala}G. Zala {\it et al}., \prb {\bf
    64}, 214204 (2001).

\bibitem{zala2} G. Zala {\it et al}.,  Phys. Rev. B {\bf 64},
  201201(R) (2001). 

\bibitem{gold2003}A. Gold, J. Phys.: Condens. Matter {\bf 15}, 217
  (2003). 

\bibitem{inter1}Y. Y. Proskuryakov {\it et al}., \prl {\bf 89},
  076406 (2002).

\bibitem{inter2}S. A. Vitkalov {\it et al}.,
  Phys. Rev. B {\bf 67}, 113310 (2003).

\bibitem{inter3}H. Noh {\it et al}., 
  \prb {\bf 68}, 165308 (2003).

\bibitem{inter4}A. A. Shashkin {\it et al}., 
   Phys. Rev. Lett {\bf 93}, 269705 (2004)

\bibitem{phonon} T. Kawamura and S. Das Sarma, Phys. Rev. B {\bf 42},
  3725 (1990).


\bibitem{drag} E. H. Hwang {\it et al}., Phys. Rev. Lett. {\bf 90},
  086801 (2003).

\bibitem{ha} M. Jonson, J. Phys. C {\bf 9}, 3055 (1976).






\end{thebibliography}
\end{document}